\documentclass[citeautoscript,floatfix,longbibliography,nofootinbib]{revtex4-2}
\usepackage{times,bm,amsmath,amsfonts,amsthm,amssymb,amsbsy,braket,hyperref,graphicx}
\usepackage[utf8]{inputenc}
\usepackage[T1]{fontenc}
\usepackage[capitalize]{cleveref}
\usepackage{microtype,physics}
\hypersetup{colorlinks=true,urlcolor=blue,linkcolor=blue,citecolor=blue}

\newcommand{\ii}{\mathrm{i}}

\begin{document}
\title{
Hidden localization transitions in canonically rotated Aubry-André models
}
\author{
Pasquale Marra%
}
\email{
pasquale.marra@keio.jp
}
%\email{pasquale.marra@yahoo.it}
%\email{pmarra@sophia.ac.jp}
\affiliation{
Department of Engineering and Applied Sciences, Sophia University, 7-1 Kioi-cho, Chiyoda-ku, Tokyo 102-8554, Japan
}
\affiliation{
Department of Physics \& Research and Education Center for Natural Sciences, Keio University, 4-1-1 Hiyoshi, Yokohama, Kanagawa, 223-8521, Japan
}
\affiliation{
Graduate School of Informatics, Nagoya University, Furo-cho, Chikusa-Ku, Nagoya, 464-8601, Japan
}

\begin{abstract}
%ABSTRACT:
Anderson localization is a phase transition between a "metallic phase", where wavefunctions are extended and delocalized in space, and an "insulating phase", where wavefunctions are completely localized.
These transitions are driven by uncorrelated or quasiperiodic disorder, e.g., in the case of the Aubry-André model.
Here, I consider a family of Hamiltonians that generalizes the Aubry-André model, obtained by replacing the position and momentum operators with an arbitrary pair of canonically conjugate operators.
These models exhibit a hidden localization transition. 
The system transitions between phases where wavefunctions are either localized or delocalized with respect to the new canonically conjugate operators, acting as an insulator or metal in this rotated space.
These canonically conjugate operators can be taken as a linear combination of position and momentum, corresponding to a "rotation" in the abstract space of canonical operators.
In this case, the hidden localization transition is signaled by the simultaneous vanishing of both the inverse participation ratio (IPR) and the normalized participation ratio (NPR) in the position and momentum space in the thermodynamic limit.
This identifies the emergence of multifractal states that are neither fully extensive nor localized on the lattice.
Hence, the states exhibit a multifractal dimension at the hidden phase transition, while remaining extended (i.e., one-dimensional) in both momentum and position everywhere else in the parameter space.
Surprisingly, I found that at the phase transition, this model Hamiltonian coincides with the lattice Hamiltonian of a massless Dirac fermion in a curved spacetime background, indicating an unexpected relation between localization transitions and analog gravity.
\end{abstract}
\date{\today}
\maketitle

\section{Introduction}
 
A paradigmatic example of localization phase transition~\cite{anderson_absence_1958,evers_anderson_2008} is provided by the Aubry-André model~\cite{aubry_analyticity_1980}, which describes particles on a one-dimensional (1D) lattice in the presence of quasiperiodic disorder.
Here, quasiperiodic disorder refers to a potential that is periodic but with a wavelength that is incommensurate with the lattice spacing, resulting in a non-repeating modulation. 
This model exhibits a transition between a "metallic phase", where all single-particle wavefunctions are spatially extended (delocalized), and an "insulating phase", where the wavefunctions are spatially confined (localized). 

This transition occurs at a finite value of the quasiperiodic disorder strength~\cite{harper_single_1955,aubry_analyticity_1980,jitomirskaya_metal-insulator_1999}.
At the phase transition, the energy spectrum becomes a fractal Cantor set with noninteger dimension~\cite{avila_solving_2006}, and coincides with the spectrum of the Harper-Hofstadter model, which describes lattice fermions on a 2D lattice in a magnetic (gauge) field~\cite{harper_single_1955,hofstadter_energy_1976}.
This corresponds to states which are neither fully localized (zero-dimensional) nor fully extended (one-dimensional) but are characterized by fractional dimensions $0<D<1$.
Such phase transitions have been observed experimentally in Bose-Einstein condensates~\cite{roati_anderson_2008}, photonic lattices~\cite{lahini_observation_2009,verbin_observation_2013,wang_localization_2020}, optical lattices~\cite{luschen_single-particle_2018}, superconducting circuits~\cite{li_observation_2023}, and acoustic crystals~\cite{ni_observation_2019}.
Localization transitions appear in both disordered and quasiperiodic systems, as well as in their nonhermitian generalizations~\cite{hatano_vortex_1997,hatano_non-hermitian_1998,hatano_localization_1996,hatano_localization_1998,luo_universality_2021,luo_unifying_2022}.
Specifically, localization transitions driven by quasiperiodicity also occur in nonhermitian Aubry-André models~\cite{longhi_metal-insulator_2019,longhi_phase_2021}, off-diagonal Aubry-André models, i.e., with nonuniform and quasiperiodic hopping amplitudes~\cite{kraus_topological_2012-2,liu_localization_2015}, and other generalizations~\cite{ganeshan_nearest_2015,wang_phase_2016,cestari_fate_2016,rossignolo_localization_2019,cookmeyer_critical_2020,cadez_machine_2023}.

The Aubry-André Hamiltonian is
\begin{equation}\label{eq:AA1D}
{\mathcal H}_\text{AA}=
J 
\left(
{e}^{\ii \hat X} + 
{e}^{-\ii \hat X} \right)+
K 
\left(
{e}^{\ii \hat Y} + 
{e}^{-\ii \hat Y} \right),
\end{equation}
where the exponentiated operators ${e}^{\ii \hat X}$ and ${e}^{\ii \hat Y}$ (known as magnetic translations in the context of the Hofstadter model) are built out of the two canonical conjugate operators $\hat X=\omega \hat x+\phi$ and $\hat Y=\hat p$ where $[\hat X,\hat Y]=\ii\omega[\hat x,\hat p]=\ii\omega$ (in natural units $\hbar=1$), and $J,K\in\mathbb{R}$.
In position basis, the Hamiltonian is\footnote{
Note that, in most of the literature~\cite{aubry_analyticity_1980,jitomirskaya_metal-insulator_1999,avila_solving_2006}, the Aubry-André Hamiltonian is written as ${\mathcal H}_\text{AA}=\sum_n J' \cos(\omega n +\phi) \ket{n}\bra{n}+ K \left( \ket{n}\bra{n+1}+\ket{n+1}\bra{n}\right)$, which is equivalent to our definition taking $J'=2J$.
}
${\mathcal H}_\text{AA}=\sum_n
2J \cos(\omega n +\phi)
\ket{n}\bra{n}+
K 
\left(
\ket{n}\bra{n+1}+
\ket{n+1}\bra{n}
\right)$.
The first term describes the quasiperiodic disorder with wavelength $2\pi/\omega$, while the second term describes the momentum contribution regularized on the lattice.
For almost all irrational values $\omega/2\pi,\phi/2\pi\in\mathbb{R}-\mathbb{Q}$ (excluding, e.g., Liouville numbers), the Aubry-André Hamiltonian exhibits a localized phase for $|J|>|K|$, with eigenstates localized in position space and delocalized in momentum space, and an extended phase for $|J|<|K|$, with eigenstates delocalized in position space and localized in momentum space, with a phase transition between the two phases at $J=\pm K$~\cite{jitomirskaya_metal-insulator_1999}.

Here, I will show that a family of equivalent Hamiltonians can be constructed by taking two arbitrary operators that satisfy $[\hat X,\hat Y]=\ii\omega$.
I will then demonstrate that these Hamiltonians exhibit a localization transition between two phases with wavefunctions respectively localized or delocalized in the basis of the eigenstates of the operator $\hat {X}$ (or equivalently $\hat {Y}$).
Because the operators $\hat X, \hat Y$ coincide with physical observables that may not be directly accessible experimentally, I refer to this transition as a \emph{hidden} localization transition.
More surprisingly, I found that this hidden phase transition can be detected not only by looking at the localization properties in the basis spanned by the eigenstates of the operator $\hat {X}$ (or $\hat {Y}$).
Indeed, the hidden phase transition is signaled by the simultaneous vanishing of the inverse participation ratio (IPR) and the normalized participation ratio (NPR) in the thermodynamic limit, and by the emergence of multifractal states with fractional dimensions $0<D<1$ in the conventional position basis.
Note that, when $\hat{X}=\frac12(\omega\hat{x}+\phi)-\hat{p}$ and $\hat{Y}=\frac12(\omega\hat{x}+\phi)+\hat{p}$, the resulting model is an off-diagonal and traceless version of the Aubry-André model, generated by a "rotation" in the abstract space of canonical operators.
Physically, this model describes a discrete lattice in which the "distances" between lattice sites (i.e., the hopping amplitudes) are modulated periodically by an incommensurate spatial frequency.
I will also demonstrate that at the transition points, this rotated Aubry-André model describes a massless Dirac fermion in curved spacetime~\cite{mcvittie_diracs_1932,mann_semiclassical_1991}, and that away from the transition points, the Hamiltonian interpolates between two different spatially-deformed metrics, which are phase-shifted relative to each other. 
This reveals an unexpected connection between localization transitions and analog gravity, i.e., quantum systems that simulate curved spacetime, such as Bose-Einstein condensates~\cite{garay_sonic_2000,lahav_realization_2010,steinhauer_observation_2016,munoz-de-nova_observation_2019}, optical metamaterials~\cite{sheng_trapping_2013,bekenstein_control_2017,zhong_controlling_2018,sheng_definite_2018,drori_observation_2019}, cold atoms in optical lattices~\cite{boada_dirac_2011}, and graphene~\cite{cortijo_a-cosmological_2007,juan_charge_2007,vozmediano_gauge_2008,de-juan_dislocations_2010,cortijo_geometrical_2012,iorio_quantum_2014,castro_symmetry_2018}.

\section{Rotated Aubry-André model} 

\begin{figure}[t]
\centering
\includegraphics[width=.8\columnwidth]{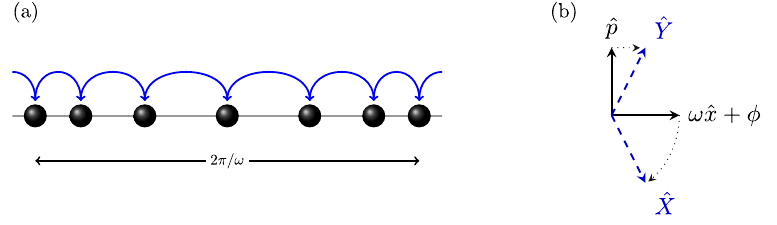}
	\caption{
(a) A sketch of the lattice Hamiltonian ${\mathcal H}_\text{RAA}$ of the rotated Aubry-André model in \cref{eq:RAAposition}, with periodically modulated hopping amplitudes, over a full period.
(b) The canonical transformation from the two canonically conjugated operators $\omega \hat x+\phi,\hat p$ to the two canonically conjugated operators $\hat X=\frac12\left(\omega\hat{x}+\phi\right)-\hat p$ and $\hat Y=\frac12\left(\omega\hat{x}+\phi\right)+\hat p$.
}
\label{fig:lattice}
\end{figure}

For a fixed spatial frequency $\omega$, any Hamiltonian ${\mathcal H}_{XY}$ with $\hat{X}$, $\hat{Y}$ being any canonical conjugate operators $[\hat{X},\hat{Y}]=\ii\omega$, is unitarily equivalent to the Aubry-André Hamiltonian as a consequence of the Stone–von Neumann theorem.
A $\pi/4$ rotation in the $X,Y$ space corresponds to the transformation $\hat{X}=(\hat{X}'-\hat{Y}')/{\sqrt{2}}$ and $\hat{Y}=(\hat{X}'+\hat{Y}')/{\sqrt{2}}$.
However, if $\hat{Y}'=\hat{p}$, such a rotation will give rise to terms $\propto e^{\frac\ii{\sqrt2}\hat{p}}$ in the Hamiltonian, which correspond to translations of $1/{\sqrt2}$ lattice sites.
However, these translations are not compatible with the lattice.
To obtain a transformation which is compatible with lattice translations, one can consider the combination of a $\pi/4$ rotation and a "rescaling" of the operators $\hat X \to \hat X/\sqrt2$, $\hat Y \to \sqrt2 \hat X$.
Consequently, taking the canonical conjugate operators obtained via a $\pi/4$ rotation and rescaling defined by
$\hat{X}=\frac12\hat{X}'-\hat{Y}'$ and $\hat{Y}=\frac12\hat{X}'+\hat{Y}'$ with $\hat{X}'=\omega\hat{x}+\phi$ and $\hat{Y}'=\hat{p}$ and using the Zassenhaus formula (see \cref{app:rotation}) yields
\begin{align}
{\mathcal H}_\text{RAA}
&=
J
e^{\frac\ii2 \left(\omega\hat x+\phi-\frac\omega2\right)} e^{-\ii \hat p}
+
K
e^{\frac\ii2 \left(\omega\hat x+\phi+\frac\omega2\right)} e^{ \ii \hat p}
+\text{h.c.}
\nonumber\\&=
\left[
2 \widetilde J
\cos\left(\frac12\left(\omega\hat{x}+\phi\right)-\frac\omega4 \right)
+
2\ii \widetilde K
\sin\left(\frac12\left(\omega\hat{x}+\phi\right)-\frac\omega4 \right)
\right]
e^{-\ii\hat{p}}
+\text{h.c.}
\nonumber\\&=
e^{-\frac\ii2\left(\omega\hat{x}+\phi\right)}
\left[
2 \widetilde J
\cos\left(\hat p-\frac\omega4\right)
+
2\ii \widetilde K
\sin\left(\hat p-\frac\omega4\right)
\right]
+\text{h.c.}
\nonumber\\&\approx
4 \widetilde J \cos\left(\frac12\left(\omega\hat{x}+\phi\right)\right) \cos\hat{p} 
%\nonumber\\&
+
4 \widetilde K \sin\left(\frac12\left(\omega\hat{x}+\phi\right)\right) \sin\hat{p}
,
\label{eq:RAA}
\end{align}
with $2\widetilde{J}=J + K$ and $2\widetilde K=J - K$ ($J=\widetilde J+\widetilde K$ and $K=\widetilde J-\widetilde K$).
This is a generalization of the model in Eq. (4) of Ref.~\onlinecite{marra_gauge_2025}.
The approximate expression on the second line holds only at small frequencies $\omega\approx0$, corresponding to a semiclassical long-wavelength limit where the commutator $[\omega\hat x,\hat p]=\ii\omega\approx0$, and it is only provided to gain some physical intuition.
For any value of the spatial frequency $\omega$, the Hamiltonian regularized in position basis $\ket{x_n}=\ket{n}$ of states localized on $x=n$ (see \cref{app:rotation}), using the shortcut notation $(a+\ii b)\otimes(c+\ii d)=a c+\ii b d$ becomes
\begin{equation}\label{eq:RAAposition}
{\mathcal H}_\text{RAA}=
\sum\limits_n
2
\left[
\left(\widetilde J+\ii \widetilde K\right)\otimes{e}^{\frac\ii2\left(\omega n +\phi-\frac\omega2\right)}
\right]
\ket{n+1}\bra{n}
+\text{h.c.}
,
\end{equation}
where $(\widetilde J+\ii \widetilde K)\otimes{e}^{\frac\ii2(\omega n +\phi-\omega/2)}
=\widetilde J\cos\left(\frac12(\omega n +\phi-\omega/2)\right)
+\ii \widetilde K\sin\left(\frac12(\omega n +\phi-\omega/2)\right)$.
In momentum basis 
\begin{equation}\label{eq:RAAmomentum}
{\mathcal H}_\text{RAA}=
\sum\limits_k
2{e}^{-\frac\ii2\phi}
\left[
\left(\widetilde J+\ii \widetilde K\right)\otimes{e}^{\ii \left(k-\frac\omega4\right)}
\right]
\ket{k+\omega/2}\bra{k}
+\text{h.c.}
,
\end{equation}
where $(\widetilde J+\ii \widetilde K)\otimes{e}^{\ii (k-\omega/4)}=\widetilde J\cos(k-\omega/4)+\ii \widetilde K\sin(k-\omega/4)$.
For periodic boundary conditions on a system of $L$ lattice sites, the spatial frequency must be commensurate with the lattice, i.e., $\omega/2=2\pi P/Q$ where $P,Q$ are two integer coprimes with $Q$ dividing $L$, such that $e^{\ii \frac12 \omega L}=1$.
The incommensurate case is obtained when $\omega/4\pi\in\mathbb{R}-\mathbb{Q}$.
\Cref{fig:lattice} shows a sketch of this lattice Hamiltonian and the corresponding canonical rotation.
Analogously to the Aubry-André model, this Hamiltonian is formally identical in position basis [\cref{eq:RAAposition}] and in momentum basis [\cref{eq:RAAmomentum}] by simply exchanging the role of position and momentum as $\frac12\omega n +\phi\leftrightarrow k$.
The rotated Aubry-André model in \cref{eq:RAA,eq:RAAposition,eq:RAAmomentum} exhibits a localization transition at $J=\pm K$, i.e., at $\widetilde J=0$ and at $\widetilde K=0$.

The canonically rotated model introduced here, where hopping amplitudes are spatially modulated and the on-site energies are identically zero, is not to be confused with another generalization of the Aubry-André model, introduced in Ref.~\onlinecite{ganeshan_nearest_2015} and experimentally realized in Ref~\onlinecite{an_interactions_2021}, where hopping amplitudes are uniform and on-site energies are spatially modulated (although with a different spatial dependence compared with the original Aubry-André model).

\section{Massless Dirac fermions in curved spacetime on a lattice} 

The spatial modulation of the hopping amplitudes in \cref{eq:RAAposition} intuitively suggests the presence of a warped or deformed spacetime metric, since the covariant derivatives in curved spacetime typically correspond to space-dependent hopping amplitudes when discretized on the lattice.
Consider a massless Dirac fermion in 1+1D curved spacetime~\cite{mcvittie_diracs_1932,mann_semiclassical_1991}
\begin{equation}\label{eq:2DDirac}
 \left[\ii\gamma^a e_a{}^\mu \partial_\mu + \frac\ii2\gamma^a \frac1{\sqrt{-g}}\partial_\mu(\sqrt{-g}\,e_a{}^\mu)\right]\psi=0,
\end{equation}
with the metric 
\begin{equation}\label{eq:metric2}
\dd s^2=\alpha(x)^2 \dd t^2 - \dd x^2,
\end{equation}
where $\psi$ is a spinor, $\gamma^\mu$ are the flat spacetime Dirac gamma matrices $\{\gamma^\mu, \gamma^\nu \}=2\eta^{\mu\nu}$ with signature $(+,-)$, and where the nonzero zweibein are $e_0{}^0=\alpha(x)^{-1}$ and $e_1{}^1=1$, and $\sqrt{-g}=\alpha(x)$. 
In the Weyl representation, $\gamma^0=\sigma_x$ and $\gamma^1=\ii\sigma_y$.
Separating time and space components,
\begin{equation}
 \ii\partial_0 \psi=
 - 
{\ii\sqrt{\alpha(x)} }
\gamma_0 \gamma^1 \partial_1 
 \left(
 \sqrt{\alpha(x)} \psi
 \right)
,
\label{eq:ContinuumHamiltonian2}
\end{equation}
which can be regularized on a discrete lattice~\cite{marra_metric-induced_2026} via
\begin{equation}\label{eq:regularization}
\partial_1\left(\sqrt{\alpha(x)}\psi\right)
\approx 
\frac1{2}\left(
\sqrt{\alpha_{n+1}}
\psi_{n+1}-
\sqrt{\alpha_{n-1}}
\psi_{n-1}
\right),
\end{equation}
with $\alpha_{n}=\alpha(n)$, which yields the lattice Hamiltonian
\begin{equation}
{\mathcal H}_\mathrm{I}=\ii
\sum\limits_n
t_n
\psi_{n+1}^\dag
 \gamma_0\gamma^1 
\psi_{n}
-
t_n
\psi_n^\dag
 \gamma_0\gamma^1 
\psi_{n+1}
,
\label{eq:DHamiltonian1}
\end{equation}
where $t_n=\frac{1}{2}\sqrt{\alpha_n\alpha_{n+1}}$.
The Hamiltonian above is equivalent up to the gauge transformation $\psi_n\to \ii^n\psi_n$ to
\begin{equation}
{\mathcal H}_\mathrm{II}=
\sum\limits_n
t_n
\left(
\psi_{n+1}^\dag
 \gamma_0\gamma^1 
\psi_{n}
+
\psi_n^\dag
 \gamma_0\gamma^1 
\psi_{n+1}
\right).
\label{eq:DHamiltonian2}
\end{equation}
Since in the Weyl representation $\gamma_0\gamma^1=-\sigma_z$, each of the Hamiltonians ${\mathcal H}_\mathrm{I}$ and ${\mathcal H}_\mathrm{II}$ decouples into two independent and identical scalar lattice Hamiltonians, one for each spinor component.

Consider the Hamiltonian ${\mathcal H}_\mathrm{I}$ with the metric 
\begin{equation}
{(\alpha_{n})}_\mathrm{I}=
\frac1{C_\mathrm{I}^{(-1)^{n}}}
\prod_{m=1}^{n-1} 
\left[\sin^2\left(\frac12(\omega m +\phi)-\frac\omega4\right)\right]
^{(-1)^{m+n+1}},
\label{eq:specialmetric1}
\end{equation}
for $n\ge1$ and with $C_\mathrm{I}$ an arbitrary number defining the boundary value $(\alpha_1)_\mathrm{I}=C_\mathrm{I}$, and giving 
\begin{equation}
t_n=\frac12\sqrt{{\alpha_n \alpha_{n+1}}}=\frac12
\left|\sin\left(\frac12(\omega n +\phi)-\frac\omega4\right)\right|.
\end{equation}
Conversely, consider the Hamiltonian ${\mathcal H}_\mathrm{II}$ with the metric 
\begin{equation}
{(\alpha_{n})}_\mathrm{II}=
\frac1{C_\mathrm{II}^{(-1)^{n}}}
\prod_{m=1}^{n-1} 
\left[\cos^2\left(\frac12(\omega m +\phi)-\frac\omega4\right)\right]
^{(-1)^{m+n+1}},
\label{eq:specialmetric2}
\end{equation}
for $n\ge1$ and with $C_\mathrm{II}$ an arbitrary number defining $(\alpha_1)_\mathrm{II}=C_\mathrm{II}$, and giving 
\begin{equation}
t_n=\frac12\sqrt{{\alpha_n \alpha_{n+1}}}=\frac12
\left|\cos\left(\frac12(\omega n +\phi)-\frac\omega4\right)\right|.
\end{equation}
Hence, consider the Hamiltonian which interpolates between ${\mathcal H}_\mathrm{I}$ and ${\mathcal H}_\mathrm{II}$ given by
\begin{equation}\label{eq:RAA'}
{\mathcal H}_\text{RAA}'=
4\widetilde J {\mathcal H}_\mathrm{II}\left[(\alpha_n)_\mathrm{II}\right]+
4\widetilde K {\mathcal H}_\mathrm{I}\left[(\alpha_n)_\mathrm{I}\right], 
\end{equation}
which becomes equivalent to the Hamiltonian ${\mathcal H}_\text{RAA}$ in \cref{eq:RAA} (when the two spinor components are decoupled), on patches where $
\sin\left(\frac12(\omega n +\phi)-\frac\omega4\right)$ and $\cos\left(\frac12(\omega n +\phi)-\frac\omega4\right)$ do not change sign.
The two metrics ${(\alpha_{n})}_\mathrm{I}$ and ${(\alpha_{n})}_\mathrm{II}$ are equivalent up to a phase shift $\delta\phi=\pi/2$.
Note that, even if the hopping amplitudes remain finite and well-defined, the metrics $(\alpha_n)_\mathrm{I}$ and $(\alpha_n)_\mathrm{II}$ diverge respectively for values of the phase such that $\omega m+\phi-\omega/2=0\mod{2\pi}$ and $\omega m+\phi-\omega/2=\pi\mod{2\pi}$ for some $m\in\mathbb{Z}$ on lattice sites where $m+n$ is even.
On a finite lattice, the values of $n,m$ are restricted to be  $1\le n,m\le L$, and therefore this issue is avoided if the phase is chosen such that $\omega (m-1/2)+\phi\neq0\mod{\pi}$ for any $1\le m\le L$.

In position basis, the Hamiltonian \cref{eq:RAA'} can be written explicitly as
\begin{equation}\label{eq:RAA'position}
{\mathcal H}_\text{RAA}'=
\sum\limits_n
2
\left[
\left(\widetilde J+\ii \widetilde K\right)\otimes{e}^{\frac\ii2(\omega n +\phi-\omega/2+\varphi_n)}
\right]
\ket{n+1}\bra{n}
+\text{h.c.}
,
\end{equation}
with the site-dependent phase shift $\varphi_n$ chosen such that the angle $0\le\frac12(\omega n +\phi-\omega/2+\varphi_n)\le\pi/2$ is the "reference angle" having the same values of trigonometric functions, up to sign, as the angle $\frac12(\omega n +\phi-\omega/2)$, giving $\cos\left(\frac12(\omega n +\phi-\omega/2+\varphi_n)\right)=|\cos\left(\frac12(\omega n +\phi-\omega/2)\right)|$ and $\sin\left(\frac12(\omega n +\phi-\omega/2+\varphi_n)\right)=|\sin\left(\frac12(\omega n +\phi-\omega/2)\right)|$.

The localization transition in \cref{eq:RAA'position} is only present when the modulation frequency $\omega$ is incommensurate to the lattice, i.e., when $\omega/4\pi$ is an irrational number.
Hence, from the point of view of analog gravity, this localization transition is an emergent lattice effect caused by the interplay between the modulation frequency and the lattice spacing.
In condensed matter, the lattice emerges as a low-energy description of a periodic potential (induced by the crystal structure in solid states or by optical lattices in cold atomic systems).
Therefore, in an analog gravity setting, the Hamiltonian \cref{eq:RAA'position} is valid in the presence of a spacetime curvature incommensurate to a periodic scalar field (e.g., electric), with the lattice parameter corresponding to the physical periodicity of the field.
Because the lattice parameter is a finite physical length, the effective metric $\Omega$ oscillates with a finite spatial frequency $\omega$, remaining smooth and differentiable.

Note also that the symmetric finite-difference scheme used in \cref{eq:regularization} inherently retains fermion doubling. 
%A more sophisticated discretization scheme (e.g., tangent/Stacey fermions) 
However, one can expect that the universal scaling properties and the location of the critical point are not affected by the fermion doubling problem, as the transition is fundamentally driven by the incommensurability of the underlying metric and scalar fields at small momenta.

\section{The hidden phase transition} 

The Hamiltonians ${\mathcal H}_\text{AA}$ and ${\mathcal H}_\text{RAA}$ are transformed one into the other by a canonical transformation $\hat{X}\to\frac12\hat{X}+\hat{Y}$, $\hat{Y}\to\frac12\hat{X}-\hat{Y}$.
This duality becomes a triality if one considers the Harper-Hofstadter Hamiltonian ${\mathcal H}_{\text{HH}}$ describing lattice fermions in a gauge field~\cite{marra_gauge_2025}.
The duality between ${\mathcal H}_\text{AA}$ and ${\mathcal H}_\text{RAA}$ mandates that also the rotated Hamiltonian ${\mathcal H}_\text{RAA}$ exhibits a localization transition at $|J|=|K|$, i.e., at $2\widetilde J=J+K=0$ and $2\widetilde K=J-K=0$, between two phases having distinct localization properties.
However, these two phases exhibit localized/delocalized modes not in the usual position basis, but in a unitarily rotated basis. 

To have an intuition about this phase transition, consider the opposite extreme regimes:
For $K=0$ (i.e., $\widetilde J=\widetilde K=J/2$), the rotated Hamiltonian is
\begin{align}
{\mathcal H}_\text{RAA}|_{K=0}
&=
J
e^{\frac\ii2 \left(\omega\hat x+\phi-\frac\omega2\right)} e^{-\ii \hat p}
+\text{h.c.}
\nonumber\\&
\approx
2 J \cos\left(\frac12\left(\omega\hat{x}+\phi\right)-\hat{p}\right) 
,
\label{eq:RAAK0}
\end{align}
with 
$(\widetilde J+\ii \widetilde K)\otimes{e}^{\frac\ii2(\omega n +\phi-\omega/2)}=\widetilde J {e}^{\frac\ii2(\omega n +\phi-\omega/2)}$
and $(\widetilde J+\ii \widetilde K)\otimes{e}^{\ii (k-\omega/4)}=\widetilde J {e}^{\ii (k-\omega/4)}$ in \cref{eq:RAAposition,eq:RAAmomentum}, and
where the approximate expression holds only in the semiclassical limit $[\omega\hat x,\hat p]=\ii\omega\allowbreak\approx0$.
In the commensurate case $\omega/4\pi=P/Q$, the Hamiltonian is diagonalized in the basis of eigenstates of the operator $\hat X=\frac12\left(\omega\hat{x}+\phi\right)-\hat p$, given by $\ket{X_n}=\sum_m{e}^{\ii\omega \frac{m^2}4 - \ii X_n m}\ket{m}$ with eigenvalues $X_n=\pi P + \phi/2+2\pi n/Q$ with $n\in\mathbb{Z}$ to satisfy the periodic boundary conditions.
Conversely, for $J=0$ (i.e., $\widetilde J=-\widetilde K=K/2$), the Hamiltonian is
\begin{align}
{\mathcal H}_\text{RAA}|_{J=0}
&=
K
e^{\frac\ii2 \left(\omega\hat x+\phi+\frac\omega2\right)} 
e^{ \ii \hat p}
+\text{h.c.}
\nonumber\\&
\approx
2 K \cos\left(\frac12\left(\omega\hat{x}+\phi \right)+\hat{p}\right) 
,
\label{eq:RAAJ0}
\end{align}
with $(\widetilde J+\ii \widetilde K)\otimes{e}^{\frac\ii2(\omega n +\phi-\omega/2)}=\widetilde J {e}^{-\frac\ii2(\omega n +\phi-\omega/2)}$
and $(\widetilde J+\ii \widetilde K)\otimes{e}^{\ii (k-\omega/4)}=\widetilde J {e}^{-\ii (k-\omega/4)}$ in \cref{eq:RAAposition,eq:RAAmomentum}, and where the approximate expression holds only in the semiclassical limit.
In the commensurate case $\omega/4\pi=P/Q$, this Hamiltonian is diagonalized in the basis of eigenstates of the operator $\hat Y=\frac12\left(\omega\hat{x}+\phi\right)+\hat p$ given by $\ket{Y_n}=\sum_m{e}^{-\ii\omega \frac{m^2}4 + \ii Y_n m}\ket{m}$ where $Y_n=\pi P + \phi/2+2\pi n/Q$ with $n\in\mathbb{Z}$ to satisfy the periodic boundary conditions.
Therefore, in the regime $|J|>|K|=0$ (i.e., $\widetilde J=\widetilde K$), the eigenstates become fully localized in the basis $\ket{X_n}$, and delocalized in the basis $\ket{Y_n}$,
while in the regime $|K|>|J|=0$ (i.e., $\widetilde J=-\widetilde K$), the eigenstates become fully localized in the basis $\ket{Y_n}$, and delocalized in the basis $\ket{X_n}$.

Right at the phase transition $J=K\neq0$ (i.e., $\widetilde K=0$ and $\widetilde J=J$), the Hamiltonian is
\begin{align}
{\mathcal H}_\text{RAA}|_{J=K}
&=
2 J
\cos\left(\frac12\left(\omega\hat{x}+\phi\right)-\frac\omega4 \right)
e^{-\ii\hat{p}}
+\text{h.c.}
\nonumber\\&\approx
4 J \cos\left(\frac12\left(\omega\hat{x}+\phi\right)\right) \cos\hat{p} 
,
\label{eq:RAAJpK}
\end{align}
which coincide with the Hamiltonian ${\mathcal H}_\mathrm{II}$ in \cref{eq:DHamiltonian2} in the metric \cref{eq:specialmetric2}.
Alternatively, at the phase transition $J=-K\neq0$ (i.e., $\widetilde J=0$ and $\widetilde K=K$),
\begin{align}
{\mathcal H}_\text{RAA}|_{J=-K}
&=
2\ii K
\sin\left(\frac12\left(\omega\hat{x}+\phi\right)-\frac\omega4 \right)
e^{-\ii\hat{p}}
+\text{h.c.}
\nonumber\\&\approx
4 K \sin\left(\frac12\left(\omega\hat{x}+\phi\right)\right) \sin\hat{p}
\label{eq:RAAJmK}
,
\end{align}
which coincide with the Hamiltonian ${\mathcal H}_\mathrm{I}$ in \cref{eq:DHamiltonian1} in the metric \cref{eq:specialmetric1}.
The Hamiltonians ${\mathcal H}_\text{RAA}|_{J=K}$ and ${\mathcal H}_\text{RAA}|_{J=-K}$ are unitarily equivalent and self-dual upon exchanging $\frac\ii2(\omega \hat x+\phi)\leftrightarrow \hat p$ (again, the approximate expression holds only in the semiclassical limit).

Let us now consider the IPR and NPR in an arbitrary basis $\ket{Z_n}$ of an operator $\hat Z$, defined as
\begin{equation}\label{eq:rotatedIPRNPR}
\text{IPR}_q^{Z}(\psi)
=
\left(
\sum_n{|\braket{\psi}{Z_n}|^{2q}}
\right)^{\tfrac1{q-1}}
,\quad
\text{NPR}_q^{Z}(\psi)
=
\frac1N
\left(
\frac1
{\sum_n{|\braket{\psi}{Z_n}|^{2q}}}
\right)^{\tfrac1{q-1}}
,
\end{equation}
which correspond to the estimated localization length $\xi^{Z}(\psi)\approx N \text{NPR}_q^{Z}(\psi)=1/\text{IPR}_q^{Z}(\psi)$ in the $Z$-space, and their average values $\langle\text{IPR}_q^{Z}\rangle=\langle\text{IPR}_q^{Z}(\psi)\rangle_\psi$, $\langle\text{NPR}_q^{Z}\rangle=\langle\text{NPR}_q^{Z}(\psi)\rangle_\psi$, and $\langle{\xi}^Z\rangle\approx N \langle\text{NPR}_q^Z\rangle\approx1/\langle\text{IPR}_q^Z\rangle$.
The usual $\text{IPR}_q^x(\psi)$, $\text{NPR}_q^x(\psi)$, and localization length $\xi^{x}(\psi)$ in the position space are recovered when $\ket{Z_n}=\ket{x_n}=\ket{n}$. 
Note that the definition of localization length used here, which can also be referred to as a "participation length", does not coincide with the localization length calculated considering the exponential decay of wave functions away from their localization centers.
This definition avoids the ambiguity of defining a single localization center in cases where the wavefunctions may be centered at various points in the lattice or exhibit multifractal structures at the transition point that lack a single well-defined exponential decay.
For these reasons, the participation length is a more physically relevant quantity to identify the hidden transition in the rotated Aubry-André model.

\begin{figure}[t]
\centering
\includegraphics[width=.9\columnwidth]{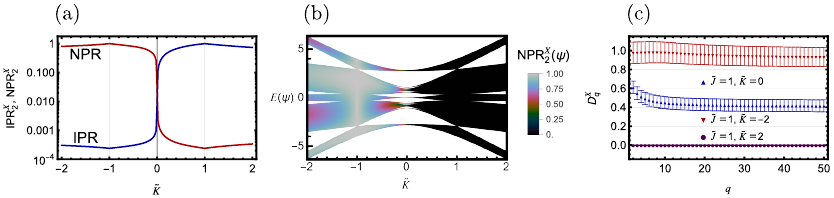}
\caption{
IPR, NPR, and multifractal dimensions of the eigenstates of the Hamiltonian ${\mathcal H}_\text{RAA}$ with modulation frequency $\omega=4\pi/\Phi_1$ calculated in the rotated basis $\ket{X_n}$. 
(a)
Average IPR and NPR in the $\ket{X_n}$ basis, and
(b)
NPR resolved in energy in the $\ket{X_n}$ basis for $2\widetilde J=J+K=1$ as a function of $2\widetilde K=J-K$.
The IPR and NPR here measure whether the eigenstates are localized or delocalized with respect to the rotated bases.
The phase transition occurs at $\widetilde K=0$ (i.e., $J=K$) separating localized states with $\langle{\text{IPR}_q}\rangle>0$ and $\langle{\text{NPR}_q}\rangle\approx0$ for $\widetilde K>0$ (i.e., $|J|>|K|$) and delocalized 
states with $\langle{\text{IPR}_q}\rangle\approx0$ and $\langle{\text{NPR}_q}\rangle>0$ for $\widetilde K<0$ (i.e., $|J|<|K|$) in the basis $\ket{X_n}$ (and vice versa for the basis $\ket{Y_n}$, see \cref{fig:FigRAA-YY}).
In the thermodynamic limit $N\to\infty$, $\langle{\text{IPR}_q}\rangle\approx0$ and $\langle{\text{NPR}_q}\rangle\approx0$ at the phase transition.
The IPR reaches its maximum $\langle{\text{IPR}_q}\rangle=1$ at $\widetilde K=1$ (i.e., $K=0$) and its minimum at $\widetilde K=-1$ (i.e., $J=0$), while the NPR reaches its minimum and maximum $\langle{\text{NPR}_q}\rangle=1$ at the same points in the basis $\ket{X_n}$ (and vice versa for the basis $\ket{Y_n}$, see \cref{fig:FigRAA-YY}).
These extrema are angular points.
(c)
Scaling dimensions as a function of the moment $q$ at $\widetilde K=\pm 1$ (i.e., $J=2$, $K=0$ or $J=0$, $K=2$) and at the transition $\widetilde K=0$ ($J=K=1$) with $\widetilde J=1$.
One has
$D_q^{X}=0$ (and $D_q^{Y}=1$, see \cref{fig:FigRAA-YY}) at $\widetilde K=1$ and in general for $|J|>|K|$,
$D_q^{X}=1$ (and $D_q^{Y}=0$) at $\widetilde K=-1$ and for $|J|<|K|$,
and multifractal dimensions $0<D_q^{X}=D_q^{Y}<1$ at the phase transition $\widetilde K=0$ ($|J|=|K|$).
}
\label{fig:FigRAA-XX}
%\end{figure}
%\begin{figure}[h]
\centering
\includegraphics[width=.9\columnwidth]{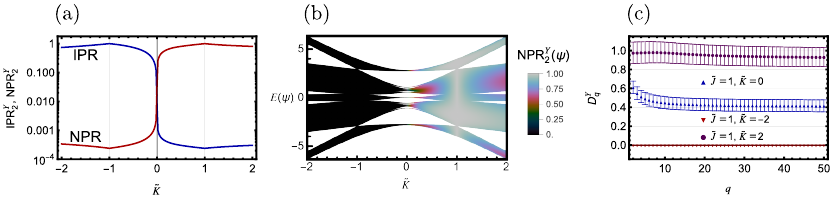}
\caption{
IPR, NPR, and fractal dimensions of the eigenstates of the Hamiltonian ${\mathcal H}_\text{RAA}$ calculated in the rotated basis $\ket{Y_n}$ as a comparison to the case of the rotated basis $\ket{X_n}$ in \cref{fig:FigRAA-XX}. 
The average IPR and NPR, and the fractal dimensions in the two bases are identical after inversion $\widetilde K\to-\widetilde K$.
}
\label{fig:FigRAA-YY}
\end{figure}

\begin{figure}[t]
\centering
\includegraphics[width=.9\columnwidth]{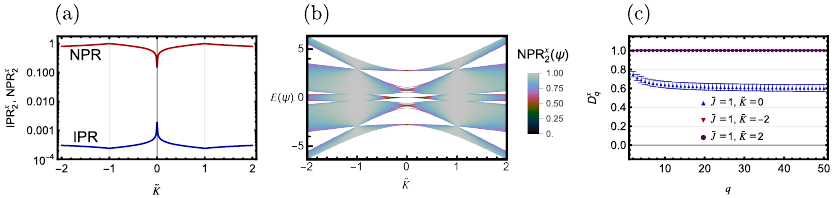}
	\caption{
IPR, NPR, and multifractal dimensions of the eigenstates of the Hamiltonian ${\mathcal H}_\text{RAA}$ with modulation frequency $\omega=4\pi/\Phi_1$ but now calculated in the usual position basis $\ket{x_n}$.
(a)
Average IPR and NPR in position basis $\ket{x_n}$, and
(b)
NPR resolved in energy in position basis $\ket{x_n}$.
The IPR and NPR in position basis measure whether the eigenstates are localized or delocalized in the conventional sense.
In position basis (and in position basis as well, see \cref{fig:FigRAA-p}), all eigenstates are delocalized for $\widetilde K\neq0$ (i.e., $|J|\neq|K|$).
The transition at $\widetilde K=0$ (i.e., $J=K$) is signaled by the minimum of the NPR and a maximum of the IPR, with both scaling to zero polynomially in the thermodynamic limit $N\to\infty$.
The IPR and NPR are symmetric with respect to the inversion $\widetilde K\to-\widetilde K$.
The NPR show maxima and the IPR minima at $\widetilde K=\pm1$ (i.e., $K=0$ and $J=0$).
These extrema are angular points.
(c) Scaling dimension, giving
$D_q^x=1$ (and $D_q^p=1$ in momentum basis, see \cref{fig:FigRAA-p}) at $\widetilde K=\pm 1$ (and for $|J|\neq|K|$) 
and multifractal dimensions $0<D_q^x=D_q^p=<1$ at the transition.
}
\label{fig:FigRAA-x}
%\end{figure}
%\begin{figure}[h]
\centering
\includegraphics[width=.9\columnwidth]{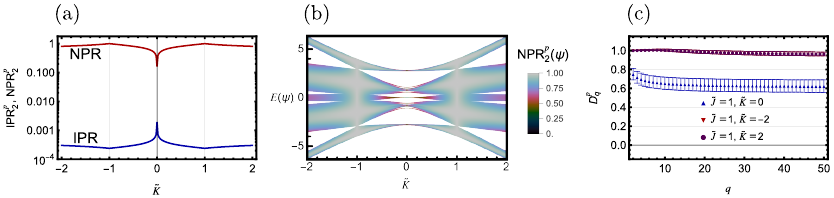}
\caption{
IPR, NPR, and fractal dimensions of the eigenstates of the Hamiltonian ${\mathcal H}_\text{RAA}$ calculated in the momentum basis $\ket{p_n}$ as a comparison to the case of the position basis $\ket{x_n}$ in \cref{fig:FigRAA-x}. 
The average IPR and NPR, and the fractal dimensions in the two bases are identical.
}
\label{fig:FigRAA-p}
\end{figure}

In the case $|J|>|K|=0$ (i.e., $\widetilde J=\widetilde K$) the Hamiltonian ${\mathcal H}_\text{RAA}|_{K=0}$ in \cref{eq:RAAK0} is diagonal in the basis of eigenstates of $\hat X=\frac12\left(\omega\hat{x}+\phi\right)-\hat p$, and hence the eigenmodes are localized on the basis $\ket{X_n}$, and extended in the basis $\ket{Y_n}$.
Hence, one can expect $\langle\text{IPR}_q^{X}\rangle\approx1$ and $\langle\text{NPR}_q^{X}\rangle\approx0$, while $\langle\text{IPR}_q^{Y}\rangle\approx0$ and $\langle\text{NPR}_q^{Y}\rangle\approx1$ 
in the case $|J|>|K|=0$ and by extension 
$\langle\text{IPR}_q^{X}\rangle>1$ and $\langle\text{NPR}_q^{X}\rangle\approx0$, while $\langle\text{IPR}_q^{Y}\rangle\approx0$ and $\langle\text{NPR}_q^{Y}\rangle>1$ in the whole phase $|J|>|K|$ (i.e., $|\widetilde K + \widetilde J|>|\widetilde K - \widetilde J|$).
In the limiting case $|J|>|K|=0$, $\langle\text{IPR}_q^{X}\rangle$ and $\langle\text{NPR}_q^{Y}\rangle$ reach their maximum values.
Exchanging the role of the bases $\ket{X_n}\leftrightarrow \ket{Y_n}$, one can expect the same behavior in the case $|K|>|J|=0$ (i.e., $\widetilde J=-\widetilde K$) and by extension in the whole phase $|J|<|K|$ (i.e., $|\widetilde K + \widetilde J|<|\widetilde K - \widetilde J|$).
At the phase transition $|J|=|K|$, the Hamiltonians ${\mathcal H}_\text{RAA}|_{J=\pm K}$ in \cref{eq:RAAJpK} and \cref{eq:RAAJmK} correspond to a critical regime at the transition between localized and extended phases, which thus leads to $\langle\text{IPR}_q^{X,Y}\rangle\approx0$ and $\langle\text{NPR}_q^{X,Y}\rangle\approx0$.

\Cref{fig:FigRAA-XX}(a), shows the IPR and NPR in the rotated basis $\ket{X_n}$ (see \cref{fig:FigRAA-YY}(a) for the basis $\ket{Y_n}$) illustrating the phase transition between the phase $\widetilde K>0$ (i.e., $J>K$) and the phase $\widetilde K<0$ (i.e., $J<K$), and with $\widetilde J=1$.
The phase $\widetilde K>0$ is localized while the phase $\widetilde K<0$ is delocalized with respect to the basis $\ket{X_n}$ (and vice versa for the basis $\ket{Y_n}$).
The IPR and NPR calculated in the two bases are identical after inversion $\widetilde K\to-\widetilde K$.

Since the transition between localized and delocalized states occurs not in position or momentum space but in a canonically rotated space, I refer to this transition as a hidden localization phase transition.
This transition is completely characterized by the behavior of the IPR and NPR in the rotated space.
In the basis $\ket{X_n}$, the eigenstates of the Hamiltonian are localized with 
$\langle\text{IPR}_q^{X}\rangle>0$ and $\langle\text{NPR}_q^{X}\rangle\approx0$
for $|J|>|K|$ (i.e., $\widetilde K>0$ in \cref{fig:FigRAA-XX}) while they are delocalized with 
$\langle\text{IPR}_q^{X}\rangle\approx0$ and $\langle\text{NPR}_q^X\rangle>0$
for $|J|<|K|$ (i.e., $\widetilde K<0$ in \cref{fig:FigRAA-XX}).
Notice that the $\langle\text{IPR}_q^{X}\rangle$ reaches its maximum value at $K=0$ (i.e., $\widetilde K=1$ in \cref{fig:FigRAA-XX}) while $\langle\text{NPR}_q^{X}\rangle$ reaches its maximum value at $J=0$ (i.e., $\widetilde K=-1$ in \cref{fig:FigRAA-XX}) according to previous considerations.
On the other hand, considering the basis $\ket{Y_n}$, the eigenstates of the Hamiltonian are delocalized with 
$\langle\text{IPR}_q^{Y}\rangle\approx0$ and $\langle\text{NPR}_q^Y\rangle>0$
for $|J|>|K|$ and localized with 
$\langle\text{IPR}_q^Y\rangle>0$ and $\langle\text{NPR}_q^Y\rangle\approx0$
for $|J|<|K|$, with $\langle\text{NPR}_q^Y\rangle$ reaching its maximum value at $K=0$ and $\langle\text{IPR}_q^Y\rangle$ reaches its maximum value at $J=0$.
The phase transition occurs at $|J|=|K|$ separating states which are localized in the basis $\ket{X_n}$ and delocalized in the basis $\ket{Y_n}$ in the phase $|J|>|K|$, and conversely delocalized in the basis $\ket{X_n}$ and localized in the basis $\ket{Y_n}$ in the opposite phase $|J|<|K|$.

However, a remnant signature of this hidden phase transition is still detectable in the usual (not rotated) position and momentum spaces.
\Cref{fig:FigRAA-x}(a) shows the IPR and NPR in position basis (see \cref{fig:FigRAA-p}(a) for the momentum basis) as a function of $2\widetilde K=J-K$, with the phase transition at $\widetilde K=0$ signaled by the dip in the NPR and a peak in the IPR in both bases.
The average IPR and NPR in both position and momentum bases show extended states for $|J|\neq|K|$, with 
$\langle\text{IPR}_q^{x}\rangle\approx0$ and $\langle\text{NPR}_q^{x}\rangle>0$.
However, at the phase transition $|J|=|K|$ (i.e., $\widetilde J=0$ or $\widetilde K=0$), the average NPR shows a dip to a minimum value corresponding to a maximum of the average IPR\@.
Both the NPR and IPR scale to zero in the thermodynamic limit at the transition.
Indeed, the IPR and NPR are symmetric with respect to the inversion $\widetilde K\to-\widetilde K$.
Intuitively, this can be understood by the fact that the two distinct phases $|J|>|K|$ and $|J|<|K|$ ($\widetilde K>0$ and $\widetilde K<0$) with states localized or delocalized in the bases $\ket{X_n}$, $\ket{Y_n}$ cannot be distinguished only by looking at the localization properties in the usual position or momentum bases, and therefore appear identical in position and momentum space.
Another feature still visible in position space is the presence of the special points $K=0$ (i.e., $\widetilde K=1$) and $J=0$ (i.e., $\widetilde K=-1$), which both exhibit a maximum in the NPR and a minimum in the IPR, and therefore appear as maximally delocalized regimes.
Hence, the eigenstates of the rotated Aubry-André Hamiltonian are always extended (i.e., one-dimensional) for $|J|\neq|K|$ but become multifractal at the hidden phase transition $|J|=|K|$ in position and momentum space, as well as in the rotated spaces spanned by the eigenstates of the canonically rotated operators.

In the usual case of the Aubry-André model, states localized in position space appear delocalized in momentum space, and vice versa.
Surprisingly, in the case of the rotated Aubry-André model, all states $|J|\neq|K|$ appear extended both in position space and in momentum space, as one can see comparing \cref{fig:FigRAA-x}(a) and \cref{fig:FigRAA-p}(a).
Indeed, in the thermodynamic limit $N\to\infty$, the IPR and NPR in momentum space in \cref{fig:FigRAA-p}(a) appear identical to those calculated in position space in \cref{fig:FigRAA-x}(a).
Intuitively, this is because position and momentum operators appear as equivalent and indistinguishable with respect to eigenstates of the Hamiltonian ${\mathcal H}_\text{RAA}$.
Loosely speaking, the eigenstates of the rotated Aubry-André model for $|J|\neq|K|$ maximize the uncertainty in both momentum and position, \emph{maximizing} Heisenberg's indeterminacy principle, while they are multifractal in both momentum and position at the transition $|J|=|K|$.

As seen above, at the phase transition, the rotated Aubry-André Hamiltonian ${\mathcal H}_\text{RAA}$ coincides with the Hamiltonian ${\mathcal H}_\text{RAA}'$ describing massless Dirac fermions in curved spacetime on a lattice, up to a space-dependent phase difference $\varphi_n$.
Obviously, the presence of this site-dependent phase shift in \cref{eq:RAA'position} introduces additional disorder such that the eigenstates of ${\mathcal H}_\text{RAA}'$ are not eigenstates of the operators $\hat X$, $\hat Y$.
\Cref{fig:FigRAAABS-XX,fig:FigRAAABS-YY} show the IPR and NPR for the Hamiltonian ${\mathcal H}_\text{RAA}'$ in the rotated bases.
Here, the IPR and NPR do not show a transition between localized and extended states.
Instead, the eigenstates appear as extended in the rotated bases for $|J|\neq|K|$, with $\langle\text{IPR}_q^{X,Y}\rangle\approx0$ and $\langle\text{NPR}_q^{X,Y}\rangle>0$.
At the phase transition $|J|=|K|$, the average NPR shows a dip to a minimum value corresponding to a maximum of the average IPR, with both scaling to zero in the thermodynamic limit.
Hence, we recover a qualitatively similar, but not identical, behavior compared to the IPR and NPR of Hamiltonian ${\mathcal H}_\text{RAA}$ in position and momentum bases in \cref{fig:FigRAA-x}.
\Cref{fig:FigRAAABS-x,fig:FigRAAABS-p} show the IPR and NPR for the Hamiltonian ${\mathcal H}_\text{RAA}'$ in position and momentum bases.
In position basis, the IPR and NPR for the Hamiltonian ${\mathcal H}_\text{RAA}'$ show an identical dependence compared with the ones obtained for the IPR and NPR of the Hamiltonian ${\mathcal H}_\text{RAA}$, as one can see comparing \cref{fig:FigRAAABS-x}(a) to \cref{fig:FigRAA-x}(a).
This corresponds to extended states for $|J|\neq|K|$ with 
$\langle\text{IPR}_q^{x}\rangle\approx0$ and $\langle\text{NPR}_q^{x}\rangle>0$
with a minimum of the NPR and a maximum of the IPR both scaling to zero in the thermodynamic limit at the transition $|J|=|K|$.
Contrary to the previous case of the Hamiltonian ${\mathcal H}_\text{RAA}$, the states are now localized in momentum space for $|J|\neq|K|$ with 
$\langle\text{IPR}_q^{p}\rangle>0$ and $\langle\text{NPR}_q^{p}\rangle\approx0$
with a maximum of the NPR and a minimum of the IPR both scaling to zero in the thermodynamic limit at the transition $|J|=|K|$, as one can see comparing \cref{fig:FigRAAABS-p}(a) to \cref{fig:FigRAA-p}(a).

Since the role of $\widetilde J$ and $\widetilde K$ in the rotated Aubry-André Hamiltonians is interchangeable, identical plots (not shown) are obtained considering the localization transition as a function of $\widetilde J$.

\section{Multifractal scaling dimension}

The scaling of the IPR and NPR with the system size is correlated with the dimensionality of the eigenstates of the Hamiltonian:
In general, 
the IPR scales as $\sim N^{-D_q^Z}$, the NPR as $\sim N^{D_q^Z-1}$, 
and the localization length as $\sim N^{D_q^Z}$,
where the dimensions $D_q^Z$ distinguishes between localized states $D_q^Z=0$ (zero-dimensional) and extended (delocalized) states $D_q^Z=1$ (one-dimensional) on a given basis $\ket{Z_n}$.
In the thermodynamic limit, this corresponds to 
a finite IPR $\sim \mathrm{const}$,
a vanishing NPR $\sim 1/N$, 
and a finite localization length 
$\langle\xi^Z\rangle\sim\mathrm{const}$ for localized states.
Conversely, it corresponds to
a vanishing IPR $\sim 1/N$,
a finite NPR $\sim \mathrm{const}$, 
and a diverging localization length 
$\langle\xi^Z\rangle\sim N$ for extended states.
At the localization transition, 
the dimensions may become fractional $0<D_q^Z<1$ and generally depend on the moment $q$, corresponding to a multifractal regime~\cite{schreiber_multifractal_1991,ohtsuki_anomalous_1997,mirlin_multifractality_2000,de-tomasi_multifractality_2020}
with the IPR and NPR both scaling to zero in the thermodynamic limit as
$\sim N^{-D_q^Z}$ and $\sim N^{D_q^Z-1}$, respectively, 
and with a localization length diverging as $\langle\xi^Z\rangle\sim N^{D_q^Z}$.
The average fractal dimensions in the basis $\ket{Z_n}$ is
\begin{equation}\label{eq:DRAA}
D_q^{Z}=
-\lim_{N\to\infty} \frac{\log\, \langle\text{IPR}_q^{Z}\rangle_N}{\log N}=
1+\lim_{N\to\infty} \frac{\log\, \langle\text{NPR}_q^{Z}\rangle_N}{\log N}
%=
%\lim_{N\to\infty} \frac{\log\, \langle\xi_q^{Z}\rangle_N}{\log N}
,
\end{equation}
where the averages $\langle\text{IPR}_q^{Z}\rangle_N$ and $\langle\text{NPR}_q^{Z}\rangle_N$ are evaluated as a function of the system size $N$.
In the case $|J|>|K|$, the eigenstates are expected to be localized (zero-dimensional) on the basis $\ket{X_n}$, giving $D_q^{X}=0$, and extended (one-dimensional) in the basis $\ket{Y_n}$, giving $D_q^{Y}=0$.
Conversely, in the case $|J|<|K|$ the eigenstates are expected to be localized (zero-dimensional) on the basis $\ket{Y_n}$, giving $D_q^{Y}=0$, and extended (one-dimensional) in the basis $\ket{X_n}$, giving $D_q^{X}=0$.
At the phase transition $|J|=|K|$, one can expect a multifractal regime with fractional dimensions $0<D_q^{X,Y}<1$, which generally depends on the moment $q$ and on the spatial frequency $\omega$.

\begin{figure}[t]
\centering
\includegraphics[width=.7\columnwidth]{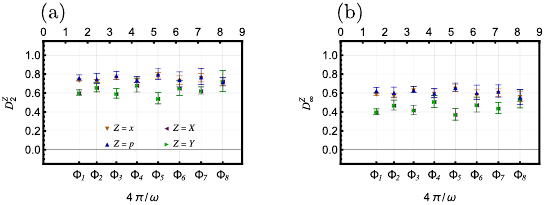}
	\caption{
The multifractal dimensions $D_q^{Z}$ of the eigenstates of the Hamiltonian ${\mathcal H}_\text{RAA}$ at the phase transition $\widetilde K=0$ ($J=K=1$) with $\widetilde J=1$, calculated in the rotated and position/momentum bases $Z=x,p,X,Y$ as a function of the spatial frequency $\omega=4\pi/\Phi_m$ with $\Phi_m=\frac{1}{2} \left(\sqrt{m^2+4}+m\right)$ the metallic ratios.
(a) Multifractal dimensions $D_2^{Z}$ for $q=2$.
(b) Multifractal dimensions $D_\infty^{Z}$ for $q\to\infty$.
In general, one has $D_q^{X}=D_q^{Y}$ and $D_q^{x}=D_q^{p}$ (up to numerical uncertainities), but 
$D_q^{X,Y}\neq D_q^{x,p}$.
}
\label{fig:FigRAA-D}
\end{figure}

Let us first look at the dimensionality of the eigenstates of the rotated Aubry-André Hamiltonian ${\mathcal H}_\text{RAA}$.
In the basis $\ket{X_n}$, the eigenstates are zero-dimensional (localized) with $D_q^{X}=0$ for $|J|>|K|$ (i.e., $\widetilde K>0$) while they are one-dimensional (delocalized) with $D_q^{X}=1$ for $|J|<|K|$ (i.e., $\widetilde K<0$) as shown in \cref{fig:FigRAA-XX}(c).
On the other hand, in the basis $\ket{Y_n}$, the eigenstates have $D_q^{Y}=1$ for $|J|>|K|$ and $D_q^{Y}=0$ for $|J|<|K|$, as shown in \cref{fig:FigRAA-YY}(c).
The phase transition occurs at $|J|=|K|$ with multifractal dimensions $0<D_q^{X}=D_q^{Y}<1$.
These two dimensions indeed coincide at the transition since the eigenstates are here symmetrical with respect to the two bases $\ket{X_n}$, $\ket{Y_n}$.
In position and momentum bases, the eigenstates are one-dimensional (extended) for $|J|\neq|K|$ with $D_q^{x,p}=1$.
However, at the phase transition $|J|=|K|$ (i.e., $\widetilde J=0$ or $\widetilde K=0$), the states have multifractal dimensions $0<D_q^{x}=D_q^{p}<1$, as shown in \cref{fig:FigRAA-D}(b).
Consequently, the peak in the IPR and the dip in the NPR both scale to zero as $\sim N^{-D_q}$ and $\sim N^{D_q-1}$, respectively.
Hence, the multifractal behavior of the rotated Aubry-André Hamiltonian at the transition is visible in the rotated bases as well as in the usual position basis.
Note that the multifractal dimensions calculated on the two rotated bases do not coincide with the one calculated in position and momentum bases $D_q^{X,Y}\neq D_q^{x,p}$.
In all cases, the multifractal dimensions at the transition scale as a function of the moment $q$ as $q D_q=q D_\infty+\mathrm{const}$.
%Hence, while the emergence of multifractality is a basis-independent signature of the hidden transition, the multifractal dimensions  $D_q^{Z}$ are not invariant under the canonical rotation, reflecting the fact that the scaling of the IPR and NPR depend on the choice of the 'ruler' used to measure the participation length.

Let us now look at the dimensionality of the eigenstates of the Hamiltonian ${\mathcal H}_\text{RAA}'$. 
In the bases $\ket{X_n}$, $\ket{Y_n}$, the eigenstates appear as one-dimensional (extended) for $|J|\neq|K|$ with $D_q^{X,Y}=1$, and become multifractal at the phase transition $|J|=|K|$ with $0<D_q^{X}=D_q^{Y}<1$, as shown in \cref{fig:FigRAAABS-XX,fig:FigRAAABS-YY}.
In position and momentum bases, the eigenstates of the Hamiltonian ${\mathcal H}_\text{RAA}'$ have dimensions $D_q^{x}=1$ and $D_q^{p}=0$ for $|J|\neq|K|$ and with multifractal dimensions $0<D_q^{x}\neq D_q^{p}<1$ at the transition $|J|=|K|$, as shown in \cref{fig:FigRAAABS-x,fig:FigRAAABS-p}.
Also, for this Hamiltonian, the multifractal dimensions calculated on the two rotated bases do not coincide with the one calculated in the position basis.
However, also note that the multifractal dimensions at the transition in position and momentum bases do not coincide in this case.

Finally, let us compare the multifractal scaling dimensions of the ${\mathcal H}_\text{RAA}$ at the phase transitions in different bases and as a function of the modulation frequency $\omega$ in \cref{fig:FigRAA-D}.
In the rotated bases $D_q^X=D_q^Y$, and in the position/momentum bases $D_q^x=D_q^p$ (up to numerical uncertainties) at all modulation frequencies.
However, the multifractal dimensions in the rotated bases and in the position/momentum bases do not coincide $D_q^{X,Y}\neq D_q^{x,p}$.
Moreover, the multifractal dimensions of the rotated Hamiltonian ${\mathcal H}_\text{RAA}$ in the rotated bases $\ket{X_n}$, $\ket{Y_n}$ appear identical to the multifractal dimensions of the Aubry-André Hamiltonian ${\mathcal H}_\text{AA}$ calculated in position and momentum bases shown in \cref{fig:FigAA-D}, i.e., $D_q^{X,Y}(\mathcal{H}_{\text{RAA}})=D_q^{x,p}(\mathcal{H}_{\text{AA}})$.
This confirms that the rotated model in the rotated bases is equivalent to the Aubry-André model in position basis.
Comparing the two rotated models ${\mathcal H}_\text{RAA}$ and ${\mathcal H}_\text{RAA}'$, one can observe that the multifractal dimensions at the transition are similar for the two Hamiltonians in position basis, i.e., $D_q^{x}(\mathcal{H}_{\text{RAA}}')\approx D_q^{x}(\mathcal{H}_{\text{RAA}})$, but not in momentum basis, since $D_q^{x}(\mathcal{H}_{\text{RAA}})\neq D_q^{p}(\mathcal{H}_{\text{RAA}})$.
Hence, the additional disorder introduced via the site-dependent phase shifts in ${\mathcal H}_\text{RAA}'$ does not affect the multifractal behavior in position space.
However, in the rotated bases, the multifractal dimensions do not coincide for the two Hamiltonians, i.e., $D_q^{X,Y}(\mathcal{H}_{\text{RAA}}')\neq D_q^{X,Y}(\mathcal{H}_{\text{RAA}})$.
Indeed, the site-dependent phase shifts make the eigenstates of the Hamiltonian incompatible with the rotated bases. 

\section{Discussion}

This work reveals the existence of hidden localization transitions in "rotated" Aubry-André models, which coincide with lattice (tight-binding) Hamiltonians with nonuniform and quasiperiodic hopping amplitudes.
This hidden phase transition corresponds to a transition between two phases where states are respectively localized or delocalized in a "rotated" space, but are always extended (i.e., one-dimensional) in the conventional position and momentum spaces, except at the transition point, where all states exhibit fractional dimensions $0<D_q<1$.
These tight-binding Hamiltonians with nonuniform hoppings describe the low-energy physics of arrays of quantum dots or atoms deposited on a surface~\cite{drost_topological_2017,palacio-morales_atomic-scale_2019} or cold atoms in optical lattices~\cite{bloch_many-body_2008,boada_dirac_2011,minar_mimicking_2015,mula_casimir_2021}, where the distances between the lattice sites are modulated in a quasiperiodic fashion.
Furthermore, this work uncovers an unexpected connection between localization and the spacetime metric: 
At the hidden phase transition, the model described in \cref{eq:RAA} becomes equivalent to the Hamiltonian of a massless Dirac fermion in a quasiperiodic spatially-deformed metric.
Indeed, several properties of the model in \cref{eq:RAA} can be inferred from its equivalence with the Aubry-André model via a canonical transformation.
Since the 1D Aubry-André model, describing fermions in periodic scalar fields on the lattice, is itself equivalent to the 2D Harper-Hofstadter model, describing fermions in gauge fields on the lattice, these equivalences form a triality between the Aubry-André, the Harper-Hofstadter, and the model in \cref{eq:RAA}.
Additional consequences of this triality are explored elsewhere~\cite{marra_gauge_2025}.

Note that, in a strictly single-particle picture, delocalized/localized wavefunctions are synonymous with metallic/insulating behavior: However, this view does not necessarily capture the physics of many-body systems, even in the noninteracting case~\cite{kohn_theory_1964}. 
A proper analysis of the localization transition would require studying charge density and its fluctuations~\cite{cookmeyer_critical_2020,hetenyi_numerical_2025,hetenyi_scaling_2024,jeon_quantum_2026}, which lies beyond the scope of this work.

In conclusion, I have introduced an off-diagonal and traceless version of the Aubry-André model, which is equivalent to the original model under a canonical transformation.
This model exhibits a hidden localization transition between phases in which the wavefunctions are respectively localized or extended in a rotated basis defined by the eigenstates of an operator that is a linear combination of position and momentum. 
I found that this hidden phase transition can be detected not only in the rotated basis, but also in the conventional position and momentum bases:
On all these bases, the transition point is signaled by the vanishing of both the inverse participation ratio (IPR) and the normalized participation ratio (NPR) in the thermodynamic limit.
This corresponds to eigenstates that have a multifractal dimension ($0<D<1$) at the transition, but remain extended (i.e., one-dimensional $D=1$) in both position and momentum everywhere else in the phase space.
Surprisingly, at the transition point, the model Hamiltonian becomes equivalent to that of a massless Dirac fermion in curved spacetime with a spatially deformed metric. 
In contrast, away from the transition, the Hamiltonian interpolates between two metrics that are phase-shifted relative to each other.
This unexpected connection between localization phase transitions and curved spacetime metric opens new avenues for the study of localization transitions, disorder, and analog gravity.

\begin{acknowledgments}
I thank Bryce Gadway, Bal\'azs Het\'enyi, Masatoshi Imada, Junmo Jeon, Alexander Mirlin, Tomi Ohtsuki, S Rahul, and Shiro Sakai for useful discussion and suggestions.
This work is partially supported by the Japan Society for the Promotion of Science (JSPS) Grant-in-Aid for Early-Career Scientists KAKENHI Grant~No.~23K13028, Grant-in-Aid for Transformative Research Areas (A) KAKENHI Grant~No.~22H05111, and Grant-in-Aid for Transformative Research Areas (B) KAKENHI Grant~No.~24H00826.
\end{acknowledgments}

\section{Data and code availability}
The code used for this study is openly available in Zenodo\cite{marra_code_2026}.

\appendix

\section{Canonical rotation\label{app:rotation}}

The Zassenhaus formula in the case that $[\hat A,\hat B]=\ii c$ with $c\in\mathbb{R}$ gives
$
e^{\hat A + \hat B} = e^{\hat A} e^{\hat B} e^{-\frac{\ii c}{2}}=e^{\hat B} e^{\hat A} e^{\frac{\ii c}{2}}
$
and 
$
e^{\ii\hat A + \ii\hat B} = e^{\ii\hat A} e^{\ii\hat B} e^{\frac{\ii c}{2}}=e^{\ii\hat B} e^{\ii\hat A} e^{-\frac{\ii c}{2}}
$.
Taking $\hat X=\hat A-\hat B$ and $\hat Y=\hat A+\hat B$ and $J,K\in\mathbb{R}$ yields
\begin{align}
{\mathcal H}_\text{RAA}
&=
J \left({e}^{\ii \hat{X}} +{e}^{-\ii \hat{X}}\right)
+
K \left({e}^{\ii \hat{Y}} +{e}^{-\ii \hat{Y}}\right)
\nonumber\\
&=
J \left({e}^{\ii \left(\hat{A}-\hat{B}\right)} + {e}^{-\ii \left(\hat{A}-\hat{B}\right)} \right)+
K \left({e}^{\ii \left(\hat{A}+\hat{B}\right)} + {e}^{-\ii \left(\hat{A}+\hat{B}\right)} \right)
\nonumber\\
&=
J \left(
e^{ \ii\hat{A}} e^{-\ii\hat{B}} e^{-\frac{\ii c}{2}}
+
e^{-\ii\hat{A}} e^{ \ii\hat{B}} e^{-\frac{\ii c}{2}}
\right)
+
K \left(
e^{ \ii\hat{A}} e^{ \ii\hat{B}} e^{ \frac{\ii c}{2}}
+
e^{-\ii\hat{A}} e^{-\ii\hat{B}} e^{ \frac{\ii c}{2}}
\right),
\end{align}
which can be written as
\begin{align}
{\mathcal H}_\text{RAA}
&=
\left(
J 
e^{-\ii\hat{A}} e^{-\frac{\ii c}{2}}
+
K 
e^{ \ii\hat{A}} e^{ \frac{\ii c}{2}}
\right)
e^{ \ii\hat{B}}
+
\left(
J 
e^{ \ii\hat{A}} e^{-\frac{\ii c}{2}}
+
K 
e^{-\ii\hat{A}} e^{ \frac{\ii c}{2}}
\right)
 e^{-\ii\hat{B}}
\nonumber\\
&=
\left[
(J + K)
\cos\left(\hat{A}+\frac{c}2\right)
-
\ii
(J - K)
\sin\left(\hat{A}+\frac{c}2\right)
\right]
e^{ \ii\hat{B}}
\nonumber\\
&+
\left[
(J + K)
\cos\left(\hat{A}-\frac{c}2\right)
+
\ii
(J - K)
\sin\left(\hat{A}-\frac{c}2\right)
\right]
e^{-\ii\hat{B}}
,
\label{eq:RAAfirstlines}
\end{align}
where the terms $\propto e^{ \ii\hat{B}}$ and $\propto e^{-\ii\hat{B}}$ are hermitian conjugates.
Alternatively, it can be written as
\begin{align}
{\mathcal H}_\text{RAA}
&=
e^{ \ii\hat{A}}
\left(
J 
e^{-\ii\hat{B}} e^{-\frac{\ii c}{2}} 
+
K 
e^{ \ii\hat{B}} e^{ \frac{\ii c}{2}}
\right)
+
e^{-\ii\hat{A}}
\left(
J 
e^{ \ii\hat{B}} e^{-\frac{\ii c}{2}} 
+
K 
e^{-\ii\hat{B}} e^{ \frac{\ii c}{2}}
\right)
\nonumber\\
&=
e^{ \ii\hat{A}}
\left[
(J + K)
\cos\left(\hat{B}+\frac{c}2\right)
-
\ii
(J - K)
\sin\left(\hat{B}+\frac{c}2\right)
\right]
\nonumber\\
&+
e^{-\ii\hat{A}}
\left[
(J + K)
\cos\left(\hat{B}-\frac{c}2\right)
+
\ii
(J - K)
\sin\left(\hat{B}-\frac{c}2\right)
\right]
,
\label{eq:RAAsecondlines}
\end{align}
where the terms $\propto e^{ \ii\hat{A}}$ and $\propto e^{-\ii\hat{A}}$ are hermitian conjugates.
In the semiclassical limit where $c\to 0$ the commutator $[\hat A,\hat B]=\ii c\approx 0$, one can write
\begin{align}
{\mathcal H}_\text{RAA}&\approx
\left[
(J+K)
\cos\hat{A}
-
\ii(J-K)
\sin\hat{A}
\right]
e^{ \ii\hat{B}}
%\nonumber\\
+
%&
\left[
(J+K)
\cos\hat{A}
+
\ii(J-K)
\sin\hat{A}
\right]
e^{-\ii\hat{B}}
\nonumber\\
&=
2(J+K)
\cos\hat{A}
\cos\hat{B}
+
2(J-K)
\sin\hat{A} 
\sin\hat{B}.
\label{eq:RAAlastlines}
\end{align}
Taking $\hat{A}=\frac12(\omega\hat{x}+\phi)$, $\hat{B}=\hat{p}$, and $c=\omega/2$ in the \cref{eq:RAAfirstlines,eq:RAAsecondlines,eq:RAAlastlines} above yields \cref{eq:RAA}.

One can verify the correctness of this result in a purely algebraic form using the Stone–von Neumann theorem.
Indeed, one can write
\begin{equation}
{\mathcal H}_\text{AA}=
J \hat V \ + K \hat T
+\text{h.c.}
,
\end{equation}
where $\hat V=e^{ \ii (\omega\hat x+\phi)}$ is a phase operator and $\hat T=e^{ \ii \hat p}$ a translation operator, with
$
\hat T \hat V= e^{\ii \omega}\hat V \hat T
$.
Analogously, one can write \cref{eq:RAA} as
\begin{align}
{\mathcal H}_\text{RAA} &=
\frac12(J+K)
\left(\hat V + \hat V^\dag\right) \hat T^\dag
+
\frac12(J-K)
\left(\hat V - \hat V^\dag\right) \hat T^\dag
+\text{h.c.}
\nonumber\\
&=
\frac12(J+K)
\left(
\hat V\hat T^\dag + \hat V^\dag\hat T^\dag
\right)
+
\frac12(J-K)
\left(
\hat V\hat T^\dag - \hat V^\dag\hat T^\dag
\right)
+\text{h.c.}
\nonumber\\
&=
\frac12(J+K)
(\hat{\widetilde V} + \hat {\widetilde T})
+
\frac12(J-K)
(\hat{\widetilde V} - \hat {\widetilde T})
 +\text{h.c.}
 \nonumber\\
&=
J
\hat{\widetilde V} 
+
K
\hat {\widetilde T}
+\text{h.c.}
,
\end{align}
where 
$\hat{\widetilde V} = \hat V\hat T^\dag=e^{\hat A-\frac{c}2}e^{-\ii \hat B}$, $\hat {\widetilde T}= \hat T \hat V=e^{ \ii \hat B}e^{\hat A-\frac{c}2}$, with $\hat V=e^{\ii A-\frac{c}2}$ and $\hat T=e^{ \ii \hat B}$.
Now one has
\begin{equation}
\hat {\widetilde T}\hat {\widetilde V} 
= (\hat{T}\hat{V})(\hat{V}\hat{T}^\dag) 
= \hat{T} \hat{V} \hat{T}^\dag
= \hat{V} \hat{T} \hat{T}^\dag e^{{\ii\omega}}
= e^{{\ii\omega}} \hat{V},
\end{equation}
and
\begin{equation}
\hat {\widetilde V} \hat {\widetilde T} 
= (\hat{V}\hat{T}^\dag)(\hat{T}\hat{V}) 
= \hat{V},
\end{equation}
which gives 
$
\hat {\widetilde T} \hat {\widetilde V}= e^{\ii \omega}\hat {\widetilde V} \hat {\widetilde T}
$.
Thus, the two sets of unitary operators $\hat { V}, \hat { T}$, and $\hat {\widetilde V}, \hat {\widetilde T}$ satisfy the same commutation relations, which corresponds to the exponentiated formulation of the canonical commutation relations (Weyl form).
Hence, for the Stone–von Neumann theorem, these two sets of operators are unitarily equivalent.
Since these two sets of operators are unitarily equivalent, there exists a unitary transformation between the two Hamiltonians ${\mathcal H}_\text{AA}$ and ${\mathcal H}_\text{RAA}$.
Note that the two sets of operators correspond to different representations of the same quantum group $\mathcal{U}_q(\mathfrak{sl}_2)$.

Regularizing with respect to the basis of states localized on $x=n$, one obtains
\begin{align}
{\mathcal H}_\text{RAA}
=
\sum\limits_n
\Bigg[
&
(J+K)
\cos\left(\frac12\left(\omega(n+1)+\phi-\frac\omega2\right) \right)
\nonumber\\
+
\ii & (J-K)
\sin\left(\frac12\left(\omega(n+1)+\phi-\frac\omega2\right) \right) 
\Bigg]
\ket{n+1}\bra{n}
+\text{h.c.}
\nonumber\\
=
\sum\limits_n
\Bigg[
&
(J+K)
\cos\left(\frac12\left(\omega n +\phi+\frac\omega2\right) \right)
\nonumber\\
+
\ii & (J-K)
\sin\left(\frac12\left(\omega n +\phi+\frac\omega2\right) \right)
\Bigg]
\ket{n+1}\bra{n}
+\text{h.c.},
\end{align}
which gives \cref{eq:RAAposition}.

\section{Details on the numerical calculations and additional figures\label{app:moreplots}}

\setcounter{figure}{0} % Resets figure counter to 0
\renewcommand{\thefigure}{B.\arabic{figure}} % Formats as A.1, A.2, etc.

\begin{figure}[t]
\centering
\includegraphics[width=.9\columnwidth]{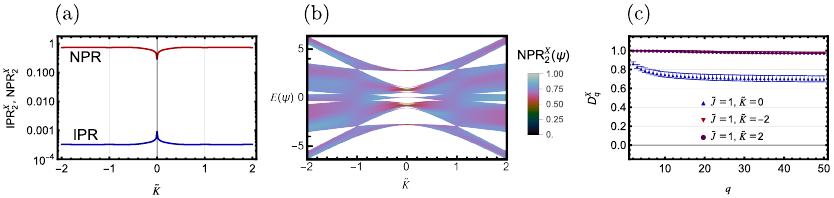}
\caption{
IPR, NPR, and fractal dimensions of the eigenstates of the Hamiltonian ${\mathcal H}_\text{RAA}'$ calculated in the rotated basis $\ket{X_n}$ as a comparison to the case of the Hamiltonian ${\mathcal H}_\text{RAA}$ in \cref{fig:FigRAA-XX}. 
The eigenstates are not diagonal in this basis in any regime.
However, a remnant signature of the phase transition is still visible in the dip of the NPR and a peak of the IPR at $\widetilde K=0$.
}
\label{fig:FigRAAABS-XX}
%\end{figure}
%\begin{figure}[h]
\centering
\includegraphics[width=.9\columnwidth]{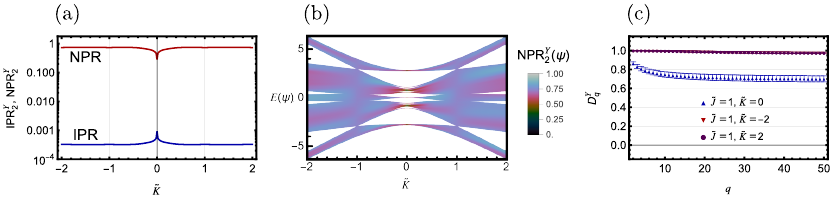}
\caption{
IPR, NPR, and fractal dimensions of the eigenstates of the Hamiltonian ${\mathcal H}_\text{RAA}'$ calculated in the rotated basis $\ket{Y_n}$, showing an identical dependence as in the case of the rotated basis $\ket{X_n}$ in \cref{fig:FigRAAABS-XX}.
}
\label{fig:FigRAAABS-YY}
\end{figure}

\Cref{fig:FigRAA-XX,fig:FigRAA-YY,fig:FigRAA-x,fig:FigRAA-p} are obtained by taking $\omega=4\pi F_{19}/F_{20}\approx4\pi/\Phi_1$ on a lattice of $N=F_{19}=4181$ ($\Phi_1$ is the golden ratio and $F_n$ the Fibonacci numbers) and $\phi/2\pi\in\mathbb{R}-\mathbb{Q}$ irrational, diagonalizing the resulting Hamiltonian with periodic boundary conditions, and calculating the IPR and NPR from the resulting energy levels wavefunctions. 
Notice that the IPR and NPR are sensitive to the degeneracy of the eigenmodes, which is unavoidable due to the time-reversal (and parity inversion) symmetry of the Hamiltonians considered.
In order to solve this issue, the calculations are done by adding a small term $\propto\sum_n e^{\ii\epsilon}\ket{n+1}\bra{n}+\text{h.c.}$ which breaks the time-reversal and parity inversion symmetry and removes the degeneracy.

The fractal dimensions are obtained by linear fitting the slope of the scaling of the logarithm of IPR and NPR $\sim D_q\log N$ (c.f., \cref{eq:DRAA}) averaged over irrational phases as a function of the system size $N=F_{n}$ in the range from $21$ to $4181$ (i.e., with $n=8,\ldots,19$), and $\omega=4\pi F_{n-1}/F_{n}\approx4\pi/\Phi_1$.
From \cref{eq:DRAA} follows that at large $N$ one has
$
{-\log\, \langle\text{IPR}_q\rangle_N}\sim\allowbreak D_q{\log N} +\mathrm{const}
$
and
$
{\log\, \langle\text{NPR}_q\rangle_N}+{\log N}\sim\allowbreak D_q{\log N} +\mathrm{const}
$.
This gives two estimates for the fractal dimensions $(D_q)_{\text{IPR}}$ and $(D_q)_{\text{NPR}}$ and the corresponding standard errors of the slope $(\sigma_q)_{\text{IPR}}$ and $(\sigma_q)_{\text{NPR}}$.
By weighting the two, we obtain the final estimate as a weighted average
$
D_q=
[
{(D_q)_{\text{IPR}}}/{(\sigma_q)_{\text{IPR}}^2}+
{(D_q)_{\text{NPR}}}/{(\sigma_q)_{\text{NPR}}^2}
]
/[
1/{(\sigma_q)_{\text{IPR}}^2}+
1/{(\sigma_q)_{\text{NPR}}^2}
]
$,
with the corresponding error given by
$
1/{\sigma_q^2}=
1/{(\sigma_q)_{\text{IPR}}^2}+
1/{(\sigma_q)_{\text{NPR}}^2}$.
Fitting the scaling dimensions for the delocalized states in the extended phase of the rotated Hamiltonian $\mathcal{H}_{\text{RAA}}$ in the rotated basis, the standard errors of the slope become zero $(\sigma_q)_{\text{IPR}}=(\sigma_q)_{\text{NPR}}=0$ and the two estimates $(D_q)_{\text{IPR}}$ and $(D_q)_{\text{NPR}}$ coincide up to numerical accuracy (corresponding to zero variance), and the same occurs for the delocalized and localized states of the Aubry-André Hamiltonian $\mathcal{H}_{\text{AA}}$ in the usual position and momentum bases.
In this cases I take $D_q=[(D_q)_{\text{IPR}}+(D_q)_{\text{NPR}}]/2$ and $\sigma_q=|(D_q)_{\text{IPR}}-(D_q)_{\text{NPR}}|/2$. 
The error bars in all plots correspond to an uncertainty $2\sigma$.
Similar plots are obtained using Hamiltonians with open boundary conditions.

The multifractal dimensions at infinite moment $D_\infty$ and the associated error $\sigma_\infty$ are then obtained by weighted linear regression on $D_q=D_{\infty}+\mathrm{const}/q$ by using $\sigma_q$ as weights.
However, the weighted linear regression assumes independent data points, which is not the case here, since the $D_q$ values are calculated from the same set of data (the wavefunctions).
Hence, the weighted linear regression may overestimate or underestimate the resulting uncertainties.
For this reason, I take the uncertainties $\sigma$ associated with $D_{\infty}$ as equal to the uncertainties associated with $D_{50}$.

\begin{figure}[t]
\centering
\includegraphics[width=.9\columnwidth]{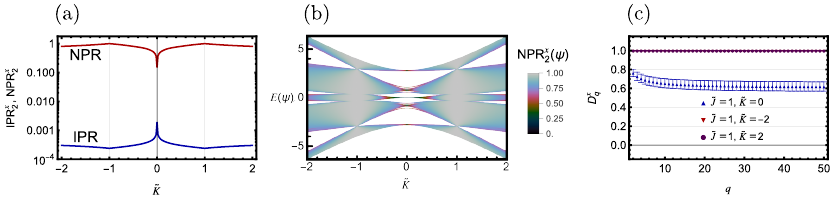}
	\caption{
IPR, NPR, and fractal dimensions of the eigenstates of the Hamiltonian ${\mathcal H}_\text{RAA}'$ calculated in the position basis $\ket{x_n}$ as a comparison to the case of the Hamiltonian ${\mathcal H}_\text{RAA}$ in \cref{fig:FigRAA-x}.
The IPR and NPR in position basis show identical behavior to the case of the Hamiltonian ${\mathcal H}_\text{RAA}$ in \cref{fig:FigRAA-x}.
}
\label{fig:FigRAAABS-x}
%\end{figure}
%\begin{figure}[h]
\centering
\includegraphics[width=.9\columnwidth]{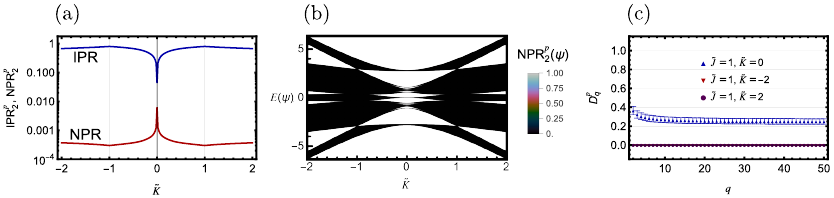}
	\caption{
IPR, NPR, and fractal dimensions of the eigenstates of the Hamiltonian ${\mathcal H}_\text{RAA}'$ calculated in the momentum basis $\ket{p_n}$ as a comparison to the case of the Hamiltonian ${\mathcal H}_\text{RAA}$ in \cref{fig:FigRAA-p}.
Interestingly, the roles of the IPR and NPR in momentum basis are reversed when compared with the case of the Hamiltonian ${\mathcal H}_\text{RAA}$ in \cref{fig:FigRAA-x}.
}
\label{fig:FigRAAABS-p}
\end{figure}

In \cref{fig:FigRAA-D}, all calculations are done as above but taking $\omega=4\pi/\Phi_m$ where 
\begin{equation}
\Phi_m=\frac{1}{2} \left(\sqrt{m^2+4} +m\right),
\end{equation} 
are the metallic ratios, such that 
\begin{equation}
\frac{\omega}{4\pi}=\frac1{\Phi_m}=
\frac{1}{2} \left(\sqrt{m^2+4}-m\right)=
\cfrac{1}{m + \cfrac{1}{m + \dots}}\,,
\end{equation}
and approximating $\omega/4\pi$ in terms of partial continued fractions $\omega/4\pi= S_{m,n-1}/S_{m,n}$.
For $m=1$ one recovers the golden ratio $\Phi_1$, with the numbers $S_{1,n}=F_{n}$ the Fibonacci numbers.
The successive approximations are obtained by restricting the partial denominators to the range from $5$ to $8658$.

\begin{figure}[t]
\centering
\includegraphics[width=.7\columnwidth]{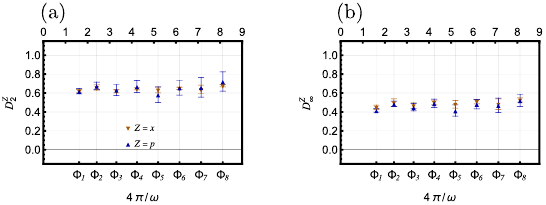}
	\caption{
The multifractal dimensions $D_q^{Z}$ as in \cref{fig:FigRAA-D}, but now calculated for the original Aubry-André Hamiltonian ${\mathcal H}_\text{AA}$ in the position and momentum bases.
In general, one has $D_q^{X}=D_q^{Y}$ and $D_q^{x}= D_q^{p}$ but $D_q^{x,p}\neq D_q^{X,Y}$ (up to numerical uncertainties).
}
\label{fig:FigAA-D}
%\end{figure}
%\begin{figure}[h]
\centering
\includegraphics[width=.7\columnwidth]{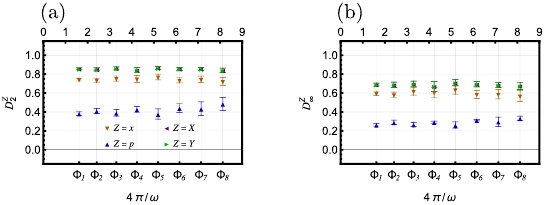}
	\caption{
The multifractal dimensions $D_q^{Z}$ as in \cref{fig:FigRAA-D}, but now calculated for the Hamiltonian ${\mathcal H}_\text{RAA}'$.
In general, one has $D_q^{X}=D_q^{Y}$ but $D_q^{x}\neq D_q^{p}$ and $D_q^{x},D_q^{p}\neq D_q^{X,Y}$ (up to numerical uncertainties) in this case.
The additional disorder (site-dependent phase shifts) in ${\mathcal H}_\text{RAA}'$ affects the multifractal behavior in the rotated bases and momentum basis, but not in position space, since $D_q(\mathcal{H}_{\text{RAA}}')\approx D_q(\mathcal{H}_{\text{RAA}})$.
}
\label{fig:FigRAAABS-D}
\end{figure}

\Cref{fig:FigRAAABS-XX,fig:FigRAAABS-YY,fig:FigRAAABS-x,fig:FigRAAABS-p} show the IPR and NPR for the Hamiltonian ${\mathcal H}_\text{RAA}'$ in the rotated and position/momentum bases for comparison with \cref{fig:FigRAA-XX,fig:FigRAA-YY,fig:FigRAA-x,fig:FigRAA-p}.
\Cref{fig:FigRAAABS-D} shows the dimensionality of the eigenstates of the rotated Hamiltonian ${\mathcal H}_\text{RAA}'$ in different bases, for comparison with \cref{fig:FigRAA-D}. 
\Cref{fig:FigAA-D} shows instead the dimensionality of the eigenstates of the original Aubry-André Hamiltonian ${\mathcal H}_\text{AA}$ in different bases.

\section{IPR, NPR, and fractal dimensions in the Aubry-André model\label{app:AAmodel}}

For completeness, I will now review here the definitions of IPR, NPR, and the fractal dimensions of the eigenstates of the Aubry-André Hamiltonian
\begin{equation}\label{eq:HH1Dposition}
{\mathcal H}_\text{AA}=\sum_n
2J \cos(\omega n +\phi)
\ket{n}\bra{n}+
K 
\left(
\ket{n}\bra{n+1}+
\ket{n+1}\bra{n}
\right).
\end{equation}
For almost all irrational values $\omega/2\pi,\phi/2\pi\in\mathbb{R}-\mathbb{Q}$, the Aubry-André Hamiltonian exhibits a localized phase for $|J|>|K|$, with localized eigenstates in position space, and an extended phase for $|J|<|K|$, with delocalized eigenstates, with a phase transition between the two phases at $J=\pm K$~\cite{jitomirskaya_metal-insulator_1999}.
Since the roles of $\hat X=\omega\hat{x} + \phi$ and $\hat Y=\hat p$ are interchangeable in \cref{eq:AA1D} upon exchanging the constants $J$ and $K$, it is clear that the statement above can be rephrased \emph{in momentum space}:
For almost all irrational values $\omega/2\pi,\phi/2\pi\in\mathbb{R}-\mathbb{Q}$, the eigenstates are localized in position space and delocalized in momentum space for $|J|>|K|$. 
In contrast, they are delocalized in position space and localized in momentum space for $|J|<|K|$, with a phase transition at $J=\pm K$.
At the phase transition $J=\pm K$, the Hamiltonian becomes self-dual under the exchange of position and momentum~\cite{aubry_analyticity_1980}.

To quantify the degree of localization, one defines $\text{IPR}_q^x$ and $\text{NPR}_q^x$ in position space via \cref{eq:rotatedIPRNPR} with $\ket{Z_n}=\ket{x_n}=\ket{n}$.
In the literature, $q=2$ is usually implied when the subscript is absent, and the global exponent $1/(q-1)$ in \cref{eq:rotatedIPRNPR} is sometimes omitted or replaced with a different one.
If all eigenstates are localized at one single lattice site, then $\text{IPR}_q^x(\psi)=1$ and $\text{NPR}_q^x(\psi)=1/N$ for each mode, giving $\langle\text{IPR}_q^x\rangle=1$ and $\langle\text{NPR}_q^x\rangle=1/N$ when averaged and $\langle\text{NPR}_q^x\rangle=0$ in the thermodynamic limit $N\to\infty$. 
Conversely, if all eigenstates are fully delocalized $\braket{\psi}{n}\sim1/\sqrt{N}$, then $\text{IPR}_q^x(\psi)\approx1/N$ and $\text{NPR}_q^x(\psi)\approx1$, giving $\langle\text{IPR}_q^x\rangle\approx1/N$ and $\langle\text{NPR}_q^x\rangle=1$ when averaged with $\langle\text{IPR}_q^x\rangle=0$ in the limit $N\to\infty$.
In general, the IPR and NPR estimate the number of lattice sites the wavefunction covers, which is $\approx N \text{NPR}_q^x(\psi)=1/\text{IPR}_q^x(\psi)$.
The eigenstates of the Aubry-André Hamiltonian have $\langle\text{IPR}_q^x\rangle>0$ and $\langle\text{NPR}_q^x\rangle\to0$ in the localized phase for $|J|>|K|$, and $\langle\text{IPR}_q^x\rangle\to0$ and $\langle\text{NPR}_q^x\rangle>0$ in the extended phase for $|J|<|K|$.
Consider the two limiting cases:
For $K=0$ in $\cref{eq:HH1Dposition}$, the kinetic term vanishes, and the eigenstates $\ket{n}$ are fully localized near the minima of the potential $\pm\cos(\omega n +\phi)$, while they are fully delocalized in momentum space.
This corresponds to a maximum of the IPR\@.
Conversely, for $J=0$ in $\cref{eq:HH1Dposition}$, the potential term vanishes, and the eigenstates are plane waves, fully delocalized in position but fully localized in momentum space.
This corresponds to a maximum of the NPR\@.

Similarly, one can define $\text{IPR}_q^p$ and $\text{NPR}_q^p$ in momentum space via \cref{eq:rotatedIPRNPR} with $\ket{Z_n}=\ket{p_n}$, where $\ket{p_n}$ are the normalized eigenstates of the momentum (plane waves $\propto e^{\ii p_n m}$ satisfying the boundary conditions.
The eigenstates of the Aubry-André Hamiltonian have $\langle\text{IPR}_q^p\rangle>0$, $\langle\text{NPR}_q^p\rangle\to0$ in the extended phase for $|J|<|K|$, and $\langle\text{IPR}_q^p\rangle\to0$, $\langle\text{NPR}_q^p\rangle>0$ in the localized phase for $|J|>|K|$.

The scaling of the IPR and NPR with the system size is correlated with the localization/delocalization of the eigenstates:
In general, $\langle\text{IPR}_q^x\rangle\sim N^{-D_q^x}$ and $\langle\text{NPR}_q^x\rangle\sim N^{D_q^x-1}$, with dimensions $D_q^x$ distinguishing between localized states $D_q^x=0$ (zero-dimensional) and extended (delocalized) states $D_q^x=1$ (one-dimensional).
At the localization transition $|J|=|K|$, the dimensions become fractional $0<D_q^x<1$ and generally depends on the moment $q$, corresponding to a multifractal regime~\cite{schreiber_multifractal_1991,ohtsuki_anomalous_1997,mirlin_multifractality_2000,de-tomasi_multifractality_2020}.

%\bibliographystyle{prsty_no_etal_titles_doi_preprint_noemph}
%\bibliography{bib}

\end{document}